\documentclass[aps,pre,showpacs,amsmath,amsfonts,amssymb,superscriptaddress]{revtex4}
\usepackage{graphics}
\usepackage[latin1]{inputenc}
\usepackage{soul}

\usepackage[pdftex]{graphicx}

\usepackage{rotating}
\usepackage[pdftex,hypertexnames=true,linktocpage=true]{hyperref}

\usepackage{color}

\usepackage{amsthm}
\usepackage{framed}
\usepackage{amssymb}
\usepackage{amsmath,mathrsfs}
\usepackage{hhline}
\usepackage{ulem}
\usepackage{fancyhdr}
\usepackage{tikz}
\everymath{\displaystyle}
\everymath{\displaystyle}

\everymath{\displaystyle}
\newcommand{\beq}{\begin{equation}}
\newcommand{\eeq}{\end{equation}}

\newcommand{\bi}{\begin{itemize}}
\newcommand{\ei}{\end{itemize}}

\newcommand{\affA}{Aix-Marseille University, Marseille, France}
\newcommand{\affB}{CNRS Centre de Physique Th\'eorique UMR7332,
13288 Marseille, France}
\newcommand{\affC}{QSTAR and INO-CNR, largo Enrico Fermi 2,
             I-50125 Firenze, Italy}

\newcommand{\affD}{School of Science and Technology, University of Camerino, I-62032 Camerino, Italy}
\newcommand{\affE}{INFN-Sezione di Perugia, Via A. Pascoli, I-06123 Perugia, Italy}

\def\RR{\mathbb{R}}

\newenvironment { abstract }
\newenvironment { acknowledgments }

\begin{document}
\title{Riemannian-geometric entropy for measuring networks complexity}

\author{Roberto Franzosi}
\email{roberto.franzosi@ino.it}
\affiliation{\affC}
\author{Domenico Felice}
\email{domenico.felice@unicam.it}
\affiliation{\affD}\affiliation{\affE}
\author{Stefano Mancini}
\email{stefano.mancini@unicam.it}
\affiliation{\affD}\affiliation{\affE}
\author{Marco Pettini}
\email{pettini@cpt.univ-mrs.fr}
\affiliation{\affA}\affiliation{\affB}

\begin{abstract}
\noindent A central issue of the science of complex systems is the quantitative characterization of complexity. In the present work we address this issue by resorting to information geometry.
Actually we propose a constructive way to associate to a - in principle any - network a differentiable object (a Riemannian manifold) whose volume is used to define an entropy. The effectiveness of the latter to measure networks complexity is successfully proved through its capability of detecting a classical phase transition  occurring in both random graphs and scale--free networks, as well as of characterizing small Exponential random graphs, Configuration Models and real networks.
\end{abstract}

\date{\today}


\maketitle

\section{Introduction}

\noindent Complex systems and phenomena are dealt with in many scientific domains. According to the domain of interest, different definitions of complexity and of the way of measuring it have been proposed and are continuously being proposed since the science of complexity is still fast growing \cite{CNet2,CNet2bis}.
The literature on this topic is so vast that any attempt at providing an exaustive bibliography would be here out of place and a very hard task. As a consequence, instead of trying to list them all,  let us notice that
 the many ways of measuring complexity belong to a restricted number of categories. In particular, the attempts at quantifying the degree of organization of a complex system often resort to some definition of an entropy function stemming from the ``archetype" represented by Shannon's information entropy \cite{shannon,shannonBis}. The latter has its precursor - at least from the point of view of physics - in Boltzmann's entropy of kinetic theory. In fact, Shannon's information entropy is equivalent to negative Boltzmann entropy, as it was proved by L. Brillouin \cite{brillouin}.

Among the different statistical-mechanical approaches to networks hitherto proposed, one of these is the class of models with hidden variables \cite{boguna}; here, the approach starts with a set of $N$ independent nodes and a general hidden variable $X$; then an undirected network is generated by: (i) assigning to each node $i$ a variable $X_i$, independently drawn from the probability $p(X)$; (ii) creating for each pair of vertices $i$ and $j$, with respective hidden variables $X_i$ and $X_j$, an undirected link with probability $p(X_i,X_j)$. So, given the independent assignment of hidden variables and links among nodes, correlated random networks are generated without neither loops nor multiple links, where the degree distribution and the correlation properties of the network are encoded in the two functions $p(X)$ and $p(X_i,X_j)\ (i,j=1,\ldots,N)$. In the present work, we consider random variables as  arbitrary hidden variables \cite{Caldarelli} sitting on the nodes, and their correlations are seen as weighted links among the nodes, again. The difference from the previous approach consists of focusing the attention on the knowledge of some parameters characterizing the hidden variables. All the informations about the system are retained in these parameters. In particular, given the information on the variances and covariances of the multiple hidden variables, a multivariate Gaussian probability distribution can be derived to describe the whole given network, by means of the Maximum Entropy Principle \cite{C09}. Thus a parameter-space is associated with any given network. This space encodes all  the  information about the structure of the associated network.  Notice that a similar way of associating  a probability distribution to a network, is that of probabilistic graphs models \cite{K12}. Actually Gaussian networks are extensively used in many applications ranging from neural networks, to wireless communication, from proteins to electronic circuits, and so on. Then, by resorting to Information Geometry \cite{AN00},  the space of the accessible values of the parameters of a given network can be endowed with the Fisher-Rao metric, so defining a Riemannian manifold. In analogy with Statistical Mechanics \cite{P07},  this manifold is the space of all the possible states of the associated network, that is, the analogous of  the phase space of a physical system. By exploiting this analogy, we may define an entropy function as the logarithm of the Riemannian volume of the manifold associated to the given network.

A first step in this direction was put forward in \cite{jmp}; in this paper, we have found that the geometric entropy associated with the Fisher-Rao metric reflects the topological features of the network: it is an increasing function of the simplices dimension. However, as it will be discussed in the following, this approach cannot be constructively applied to networks having more than a few nodes. 

Then, in \cite{EPL} a new metric - obtained by a suitable ``deformation" of the standard Fisher-Rao metric of information geometry - was proposed which allows to constructively lift the properties of any given network to the geometric structure of a manifold.  In this way, we associated a differentiable system (Riemannian manifold) to a discrete system (network) through the description of network by a set of probability distribution functions. Among the wide use of probabilistic methods in literature it is worth mentioning the random walk method \cite{Noh}. Here the Green function, meaning the transition amplitude from one vertex to another by accounting for all possible walks, gives rise to a metric \cite{Blachere}, thus allowing as well for a geometric approach. However the main difference is that by considering random walk the (transition) probability is given through the adjacency matrix, while in our case probabilities are given through Gaussian joint distribution of random variables sitting on nodes of the network. In addition, as it is clearly shown in \cite{EPL},  such a geometric entropy is able to detect  the classical transition in random graphs predicted by the Erd\"{o}s-R\'enyi theorem \cite{ER,Lucz}.

Here we want to propose that such an entropy qualifies as a networks complexity measure. To this end we deepen the study about the random graphs model of \cite{EPL} and then validate our measure on complex networks. 

The layout of the paper is as follows. For the sake of self consistency and readability of the paper, the mathematical framework is reported in Sections II and III. In particular, in Section II we briefly recall the 
relation between Gaussian statistical model and underlying network
putting forward the metric structure of the associated manifold.
While in Section III we present the geometric measure of complexity as the 
logarithm of the Riemannian volume of the manifold. To the end of validating this complexity measure, we check it against both theoretical and empirical networks, spreading from Random Graphs to real networks by stepwise increasing ``the degree of complexity''. In Section IV we report a strong validation point of our  complexity measure of networks,  a first account of which was given in  \cite{EPL}.  This first validation was obtained by computing our geometric entropy for the Erd\"os-R\'enyi phase transition in Random Graphs, that is by checking how it performs against a rigorous analytical result on networks. In Section V we extend our investigation on different network models. We start considering theoretical models of Exponential Random Graphs and Configuration Models which are well-known and studied in the literature.  Next, we show that our geometric entropy is also able to detect the emergence of a giant component as predicted by the Molloy and Reed criterium for scale--free networks. Then, we apply  our complexity measure to real networks and compare the outcomes so obtained to the results that have been already reported in the literature on these same systems using other complexity measures. Section VI is devoted to possible future developments concerning the possibility to predict the stability of a network system. Conclusions are drawn in Section VII.


\section{Information geometric model}

Usually in mathematics in order to get information on a geometric object one endows it with a superstructure (e.g. bundles over manifolds, coverings over topological spaces, and so on).
Likewise we endow a network with a statistical Riemannian manifold. This can be obtained basically via two steps; first by understanding a network as an undirected graph without loops on the nodes, and account for links (weighted edges) between nodes expressed by the adjacency matrix $A$ by means of correlations. Then, by associating  some random variables with the vertices of a network, one can resort to the methods of Information Geometry \cite{AN00} to associate a statistical Riemannian manifold with the network. 

So, let us consider a set of $n$ real-valued random variables $X_1,\ldots,X_n$ distributed according to 
a multivariate Gaussian probability distribution 
(assumed for the sake of simplicity of zero mean)
\begin{equation}
 p(x;\theta)=\frac{1}{\sqrt{(2\pi)^n\det C}}\exp\left[-\frac 1 2 x^t {C}^{-1}x\right],
\label{PxT}
\end{equation}
where $x^t=(x_1,\ldots,x_n)\in\RR^n $ with $t$ denoting the transposition. Furthermore, $\theta^t=(\theta^1,\ldots\theta^m)$ are the real valued parameters characterizing the above probability distribution function, namely the entries of the covariance matrix $C$. As a consequence $m=n(n+1)/2$.  

Next consider the family $\cal P$ of such probability distributions 
\begin{equation*}
{\cal P}=\{p_\theta=p(x;\theta)|\theta^{t}=(\theta^1,\ldots\theta^m)\in\Theta\},
\end{equation*}
where $\Theta\subseteq\RR^m$. Upon requiring the mapping $\theta\rightarrow p_\theta$ to be injective, $\cal P$ becomes an $m$-dimensional statistical model on $\RR^n$. The open set $\Theta$ results defined as follows
\begin{equation}\label{parameterspace}
\Theta=\{\theta\in \RR^m|C(\theta)>0 \},
\end{equation}
and we refer to it as the parameter space of the statistical model $\cal P$.

Since any element $p(x;\theta)\in \mathcal{P}$ is univocally characterized by the parameter vector $\theta$, it follows that the mapping $\varphi:{\cal P}\rightarrow \Theta$ defined by $\varphi(p_\theta)=\theta$ is a coordinate chart. So, $\varphi=[\theta^i]$ can be considered as a local coordinate system for $\cal P$. Then $\cal P$ can be turned into a $C^\infty$ differentiable manifold by
assuming parametrizations that are $C^\infty$ \cite{AN00}.

Given an $m$-dimensional statistical model ${\cal P}=\{p_\theta|\theta\in\Theta\}$ its Fisher information matrix 
in $\theta$ is the $m\times m$ matrix $G(\theta)=[g_{\mu\nu}]$, whose entries are defined by
\begin{equation}
g_{\mu\nu}(\theta):=\int_{\RR^n} dx \;p(x;\theta)\partial_\mu\log p(x;\theta)\partial_\nu\log p(x;\theta),
\label{gFR}
\end{equation}
with $\partial_\mu\equiv\frac{\partial}{\partial\theta^\mu}$. The matrix $G(\theta)$ results symmetric, 
positive definite and provides a Riemannian metric for the parameter space $\Theta$ \cite{AN00}. 

For our case the integral in Eq. \eqref{gFR} is Gaussian and can be computed as
\begin{eqnarray}
&&\frac{1}{\sqrt{(2\pi)^n\det C}}\int dx f_{\mu\nu}(x) \exp\left[-\frac 1 2 x^t C^{-1} x\right]\nonumber\\
&&=\exp\left[\frac 1 2 \sum_{i,j=1}^n 
c_{ij}\frac{\partial}{\partial x_i}\frac{\partial}{\partial x_j}\right]f_{\mu\nu} |_{x=0},
\label{Gint}
\end{eqnarray}
where 
\begin{equation}\label{f} 
f_{\mu\nu}:=\partial_\mu \log[p(x;\theta)]\ \partial_\nu \log[p(x;\theta)],
\end{equation}
and the exponential stands for a power series expansion over its argument (the differential operator). 
The derivative of the logarithm reads
\begin{eqnarray}\label{logder}
&&\partial_\mu \log[p(x;\theta)]=-\frac 1 2\Bigg[\frac{\partial_\mu(\det C)}{\det C}
+\sum_{\alpha,\beta=1}^n \partial_\mu(c_{\alpha\beta}^{-1})x_\alpha x_\beta\Bigg],\nonumber\\
\end{eqnarray}
where $c_{\alpha\beta}^{-1}$ denotes the entries of the inverse of the covariance matrix $C$.

The computational complexity of the metric components in Eq. \eqref{gFR} can be readily estimate. Indeed, the well-known formulae 
\begin{eqnarray*}
\partial_\mu C^{-1}(\theta)&=&C^{-1}(\theta)\big(\partial_\mu C(\theta)\big)C^{-1}(\theta),\\
\partial_\mu(\det C(\theta))&=&\det C(\theta) \;\mbox{Tr}(C(\theta)\,\partial_\mu(C(\theta))),
\end{eqnarray*}
 require the calculation of $n(n+1)$ derivatives, with respect to the variables $\theta\in\Theta$, in order to work out the derivative of the logarithm in \eqref{logder}. Then, to obtain the function $f_{\mu\nu}$ in \eqref{f}, we have to calculate $O(n^4)$ derivatives. With growing $n$ this becomes a daunting task, even when afforded numerically. 


\subsection{An alternative to the Fisher-Rao metric}

In order to overcome the difficulty of computing the components of the Fisher-Rao metric, we 
follow \cite{EPL} and define a (pseudo)-Riemannian metric on the parameter space $\Theta$  which account as well for the network structure given by the adjacency matrix $A$.

 To this end we consider first a trivial network with null adjacency matrix that is associated with a set of  $n$ independent Gaussian random variables $X_i$. Notice that in this particular case, the covariance matrix in \eqref{PxT} is a diagonal matrix with entries given by $\theta^i:=\mathbb{E}(X_i^2)$. Let us denote this matrix as $C_0(\theta)$. So, employing Eqs. \eqref{parameterspace} and \eqref{gFR},  a statistical Riemannian manifold $\mathcal{M}=(\Theta,g)$, with 
 \begin{equation}\label{diagonal}
 \Theta=\{\theta\equiv(\theta^1,\ldots\theta^n)|\theta^i>0\},\quad g=\frac{1}{2}\sum_{i=1}^n\Big(\frac{1}{\theta^i}\Big)^2d\theta^i\otimes d\theta^i \ ,
 \end{equation}
is associated to the bare network.

Let us remark that the entries $g_{ii}$ of the metric $g$ in \eqref{diagonal}, worked out in \cite{jmp},
depend on the entries of the matrix $C_0(\theta)$. In fact, the $ii$ entries of the inverse matrix of $C_0(\theta)$ are given by $c_{ii}^{-1}=\frac{1}{\theta^i}$. Then, from \eqref{diagonal} it is evident that $g_{ii}=\frac{1}{2}(c_{ii}^{-1})^2$. Inspired by
this functional form of $g$, we associate a (pseudo)-Riemannian manifold to any network $\cal X$ with non vanishing  adjacency matrix $A$ by ``deforming" the manifold ${\cal M}$ in \eqref{diagonal} via the map
 $\psi_{C_0}:\mbox{A}(n,\RR)\rightarrow\mbox{GL}(n,\RR)$ defined by
\begin{equation}\label{psi} 
\psi_{C_0(\theta)}(A):=C_0(\theta)+A.
\end{equation}
By $\mbox{A}(n,\RR)$ we denote the set of the symmetric $n\times n$ matrices over $\RR$  with vanishing diagonal elements that can represent  any simple undirected graph. 
Therefore, the  manifold associated to a network $\cal X$, with adjacency matrix $A$, is  $\widetilde{\cal M}=(\widetilde{\Theta},\widetilde{g})$. Here it is   
 \begin{equation}
 \label{varyspace}
 \widetilde{\Theta}:=\{\theta\in\Theta\ \vert\ \psi_{C_0(\theta)}(A)\ \mbox{is non-degenerate}\},
 \end{equation}
 and  $\widetilde{g}=\sum_{\mu\nu}\widetilde{g}_{\mu\nu}d\theta^{\mu}\otimes d\theta^{\nu}$ with components
\begin{equation}\label{gvary}
\widetilde{g}_{\mu\nu}=\frac{1}{2}(\psi_{C_0(\theta)}(A)^{-1}_{\mu\nu})^2,
\end{equation}
where $\psi_{C_0(\theta)}(A)^{-1}_{\mu\nu}$ are the entries of the inverse of the matrix $\psi_{C_0(\theta)}(A)$.


\section{A measure of networks complexity}

We now define a statistical measure of the complexity of a network $\cal X$ with adjacency matrix $A$ and associated manifold $\widetilde{\cal M}=(\widetilde{\Theta},\widetilde{g})$ as
\begin{equation}
{\cal S}:= \ln {\cal V}(A),
\label{entropy}
\end{equation}
where ${\cal V}(A)$ is the volume of $\widetilde{\cal M}$ evaluated from the element 
\begin{equation}
\nu_g=\sqrt{|\det \widetilde{g}(\theta)|}\;d\theta^1\wedge\ldots\wedge d\theta^n\ .
\label{volumelement}
\end{equation}
Notice, however, that in such a way ${\cal V}(A)$ results ill-defined.  
In fact, the set $\widetilde{\Theta}$ in Eq.\eqref{varyspace} is not compact because the variables $\theta^i$ are unbound from above. Furthermore, from Eq.\eqref{gvary}, $\det \widetilde{g}(\theta)$ diverges
since $\det \psi_{C_0(\theta)}(A)$ approaches zero for some $\theta^i$. 

Thus, as it commonly happens \cite{Leibb}, we regularize the volume as follows
\begin{equation}
{\cal V}(A):=\int_{\widetilde{\Theta}}\Upsilon(\psi_{C_0(\theta)}(A))\;\nu_g,
\label{reg}
\end{equation}
where   $\Upsilon(\psi_{C_0(\theta)}(A))$ is any suitable  "infrared" and "ultraviolet" regularizing function,
i.e. providing a kind of compactification of the parameter space and excluding
the contributions of $\theta^i$ making $\det \widetilde{g}(\theta)$ divergent. Theoretically, a regularizing function $\Upsilon(\psi_{C_0(\theta)}(A))$ might be devised by taking into account particular structures of the integration set \eqref{varyspace} and the functional relation \eqref{gvary}. In practice, a very suitable function has been built in Ref.\cite{jmp}; it reads as,
\begin{align}\label{regjmp}
\Upsilon(C(\theta))=\log\left[1+\det \left(C(\theta)\right)^n\right]\ e^{-\mbox{tr}C(\theta)},
\end{align}
when the covariance matrix $C(\theta)$ has only $1$ or $0$ as off diagonal entries. Here, the logarithm hales contributions of $\theta^i$ that make $\det C$ diverge at the lower bound of the parameter space; while, the exponential fixes the problem of non-compact integration space. In this paper, we tackle networks with weighted links among the nodes. Still the functional type of the regularizing function is like in \eqref{regjmp}.

The definition \eqref{entropy} is inspired by the microcanonical definition of entropy in statistical mechanics
that is proportional to the logarithm of the volume of the Riemannian manifold associated with the underlying dynamics \cite{P07}.

Of course we need to validate the proposed measure of network  complexity defined 
in Eq.\eqref{entropy}. Though in principle any measure of complexity is admissible, we may wonder how to assess its effectiveness. 
A first step is to check a complexity measure against a system which makes a clear jump of complexity as some parameter is varied. In physics a paradigmatic situation is offered by phase transitions (a snowflake is intuitively more complex than a drop of water). Applied to networks this leads us to consider the classical Erd\"{o}s-R\'enyi phase transition in random graphs \cite{ER,Lucz}.
Then, moving on from random graphs, more complex networks can be considered and the proposed measure of 
complexity compared with other known measures.
These will be the subjects of the following Sections.


\section{The  Erd\"{o}s-R\'enyi phase transition}

One of the basic models of random graphs is the \textit{uniform random graph} $\mathbb{G}(n,k)$. This is devised by choosing  with uniform probability a graph from the set of all the graphs having $n$ vertices and $k$ edges, with $k$ a non negative integer. We can think of $\mathbb{G}(n,k)$ as a process evolving by adding the edges one at a time. When $k$ has the same order of magnitude of $n$, the evolution of $\mathbb{G}(n,k)$ from $k=0$ to $k=\binom{n}{2}$ yields, according to Erd\"{o}s-R\'enyi theorem \cite{ER}, a \textit{ phase transition}, revealing itself  in a rapid growth with $k$ of the size of the largest component (number of vertices fully connected by edges). Specifically, the structure of $\mathbb{G}(n,k)$ when the expected degree of each of its vertices is close to $1$, i.e. $k\sim n/2$, shows a jump: the order of magnitude of the size  of the largest component of $\mathbb{G}(n,k)$ rapidly grows, asymptotically almost surely (a.a.s.), from $\log n$ to $n$, if $k$ has the same order of magnitude of $n$ .
In fact, if $k<n/2$, as the process evolves, the components of $\mathbb{G}(n,k)$ [the largest of them being a.a.s. of size $O(\log n)$] merge mainly by attaching small trees; thus they grow slowly and quite smoothly  \cite{Lucz}. Nonetheless, at the same point of the process, the largest components become so large that it is likely for a new edge to connect two of them. Thus, fairly quickly, all the largest components of $\mathbb{G}(n,k)$ merge into one giant component, much larger than any of the remaining ones \cite{Lucz}.  
It is worth noticing that this process represents the mean-field case of percolation \cite{PERC}.

We numerically compute ${\cal S}(k)$, the geometric entropy in Eq.\eqref{entropy} vs $k$ for a fixed $n$, in order to investigate its sensitivity to the appearance of the giant component during the evolution of the random graph model
$\mathbb{G}(n,k)$.

It is worth mentioning that a Gibbs entropy has been defined for the statistical set of Random Graphs \cite{bogacz} as
 
\begin{equation}
S:=\ln \frac{1}{n!} \binom{\binom{n}{2}}{k}.
\label{entrobianco}
\end{equation}
Following up, a research line has been pursued to characterize other classes of random graphs, like scale-free or fixed-degree sequence \cite{bianco}. However, the entropy \ref{entrobianco} (as function of $k/n$) is not able to detect the Erdos-R\'enyi phase transition.

In practice we have considered four different numbers of vertices:  $n=25, 50, 100, 200$. Notice that the magnitude of $n$ is not important, what matters is the $n$-dependence, the so-called finite size scaling, of the relevant observables.
The magnitude of $n$ simply determines the dimension of the manifold ${\widetilde{\cal M}}$. For any fixed $n$ we have considered the number of links $k$, to be $k=0,1,\ldots,n(n-1)/2$. Then, for a any pair $(n,k)$ we have randomly generated a set of $k$ entries $(i,j)$, with $i<j$, of the non-vanishing adjacency matrix elements $A_{ij}$.

In this way, since the covariance  matrix $C$ is functionally assigned, we have gotten $\psi_{C}(A)$ of  Eq. \eqref{psi} and finally the metric $\widetilde{g}$ of Eq.\eqref{gvary}. Next, having determined ${\widetilde{\cal M}}=({\widetilde{\Theta}},\widetilde{g})$, we computed the volume ${\cal V}(A)$ in Eq.\eqref{reg} and the entropy ${\cal S}$ of Eq.\eqref{entropy}. The volume regularization is performed in two steps. First by restricting the manifold support ${\widetilde{\Theta}}\subset \mathbb{R}^n$ to an hypercube.
Inside it we generated a Markov chain to perform a Monte Carlo estimation of the average $\langle\sqrt{\det \widetilde{g}}  \rangle =\int \sqrt{\det \widetilde{g}}\;d\theta^1\wedge\ldots\wedge d\theta^n{ \bigg/} \int \;d\theta^1\wedge\ldots\wedge d\theta^n$. The number of considered random configurations ranges between $10^4$ and $10^6$. As second step of the regularization we have excluded those points where the value of $\sqrt{\det \widetilde{g}}$ exceeds $10^{308}$ (the numerical overflow limit of the computers used). 
Then, for any given pair $(n,k)$ this computational procedure is repeated $10^3$ times, each time considering a different randomly generated realization of the adjacency matrix $A$. Thus, the final values of the entropy  
${\cal S}$ are obtained as averages over $10^3$ different manifolds ${\widetilde{\cal M}}$, namely
\begin{eqnarray}
&&{\widetilde{\cal S}}(k) := \frac{1}{n}\langle ({\cal S}(k) - {\cal S}(0))\rangle\nonumber \\
&&= \frac{1}{n} \left\langle\ln \frac{\int \sqrt{\det \widetilde{g}}\;d\theta^1\wedge\ldots\wedge d\theta^n}{ \int \sqrt{\det { g}} \;d\theta^1\wedge\ldots\wedge d\theta^n}\right\rangle
\label{collapseplot}
\end{eqnarray}
where $g$ is the metric corresponding to the null adjacency matrix.

\begin{figure}[h!] \centering
\includegraphics[width=7cm,height=6cm,scale=1.]{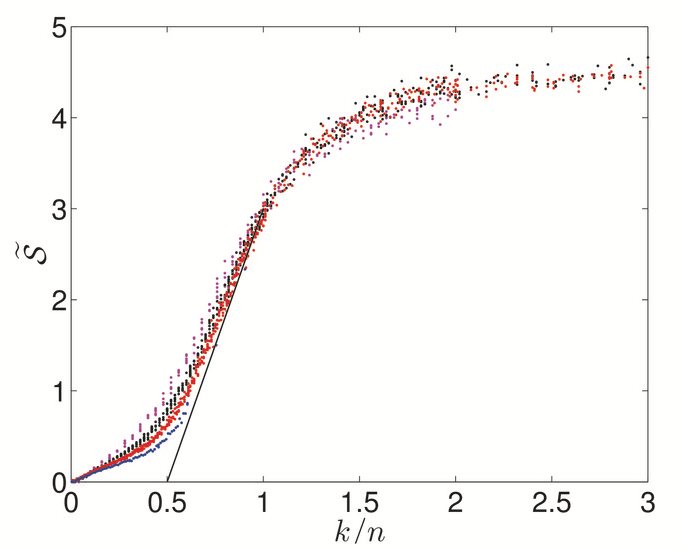}
\caption{(Color online) Values of the entropy $\widetilde{\cal S}(k/n)$ of  $\mathbb{G}(25,k)$ (magenta points), $\mathbb{G}(50,k)$ (black points), $\mathbb{G}(100,k)$  (red points) and $\mathbb{G}(200,k)$  (blue points) networks as a function of the number $k$ of randomly chosen links of weights equal to $r=0.2$. The black solid line is a guide to the eye coming from a linear fitting of a linear-logarithmic presentation of the data.}
\label{bifurcation}
\end{figure}

In Figure \ref{bifurcation} we report the behavior of $\widetilde{\cal S}(k/n)$
of the case of equal weights { $A_{ij}=r$} for all the $k$ non-vanishing links.
This is what in the context of statistical mechanics is known as a \textit{collapse  plot} of the results obtained at different $n$-values. It shows a typical phenomenon arising in numerical investigations of second order phase transitions: likewise finite-size effects observed for the order parameter, what asymptotically would be a sharp bifurcation is rounded at finite $n$. However, the larger $n$ is, the more pronounced the "knee" of ${\widetilde{\cal S}(k/n)}$ becomes. This is agreement  with an $n$-asymptotic bifurcation at $k/n = 0.5$ (black solid line) where the Erd\"{o}s-R\'enyi phase transition takes place. 

\begin{figure}[h!] \centering
\includegraphics[width=7cm,height=6cm,scale=1.]{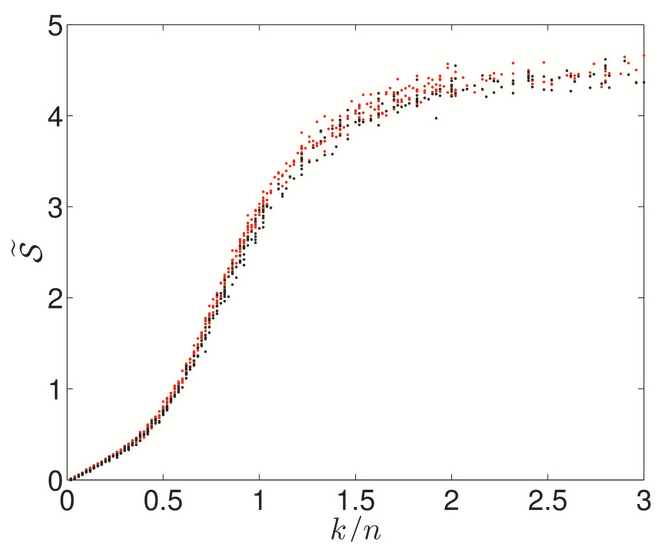}
\caption{(Color online) Values of the entropy $\widetilde{\cal S}(k/n)$ of  $\mathbb{G}(50,k)$ networks as a function of the number $k$ of randomly chosen links of weight equal to $r=0.2$ (red points), $r=0.2+\omega$  with $\omega$ a gaussian random of zero mean and variance $0.1$ (black points).}
\label{rndeffects}
\end{figure}

At present, this beautiful and unambiguous result (presented also in \cite{EPL}) lends credit to our proposed measure of networks complexity. To enforce it we a stability check would be in order.
Then, in Figure \ref{rndeffects} we report the outcomes of $\mathbb{G}(50, k)$ having chosen at random the values of the non-vanishing entries $A_{ij}$ of the adjacency matrix, that is, $A_{ij}=0.2+\omega$ where $\omega$ is a random variable of zero mean and variance equal to $0.1$. Of course negative values of the $A_{ij}$ are excluded. The comparison with the results obtained with $A_{ij}=0.2$ confirms the robustness of the entropy defined in Eq.\eqref{entropy}.

\section{Beyond random graphs}

Here we go beyond the random graph model and apply the proposed measure complexity defined 
in Eq.\eqref{entropy} to complex networks with the aim of comparing our results with other already known.

\subsection{Small Exponential Random Graphs}

The general idea that a system is complex when it does not coincide with the ``summation" of all its parts, has been formalized in \cite{jostchaos} within the framework of Information Geometry. With this approach, a hierarchy of exponential families is provided, which is widely studied in information geometry \cite{Amari}, modeling networks of progressively increasing order of the interactions between their parts. The model known as Exponential Random Graphs (ERG) is the distribution over a specified set of graphs that maximizes the Gibbs entropy under some suitable constraints;  more precisely, let us suppose to be given a collection $\left\{\xi_i\right\}_{i=1,\ldots,r}$ of graph observables (such as nodes and edges) measured in empirical observations of some real-world network, or of any other network as well. In addition, assume that
we have an estimate $\langle \xi_i\rangle$ of the expectation value of each observable. Consider now a graph $G$, and $P(G)$ the probability of that graph within a given ensemble ${\cal G}$. Then we wonder which is the best choice for $P(G)$ so that the expectation value of each of our graph observables $\left\{\xi_i\right\}$ within that distribution is equal to its observed value. The answer is obtained by maximizing Gibbs' entropy
\begin{equation*}
S=-\sum_{G\in{\cal G}}P(G)\ln P(G)
\end{equation*}
  
under the constraints $\sum_{G\in{\cal G}}P(G)\xi_i(G)=\langle \xi_i\rangle$ and $\sum_{G\in{\cal G}}P(G)=1$, where $\xi_i(G)$ is the value of $\xi_i$ for the graph $G$. This procedure leads to
\begin{equation}
P(G)=\frac{e^{-{\cal H}(G)}}{Z}, 
\end{equation}
where ${\cal H}(G)=\sum_i \lambda_i \xi_i(G)$ is the graph Hamiltonian and $Z=\sum_{G\in{\cal G}} e^{-{\cal H}(G)}$ is the partition function. Here, $\lambda_i$s are Lagrangian multipliers. This equation defines the Exponential Random Graph model \cite{park}.

This model has been employed to quantify the degree of interaction of all the parts of a given system \cite{Olbrich}. Still in Ref.\cite{Olbrich}, simple exponential random graphs are considered in order to describe ``typical'' graphs, i.e. the graphs that are most probable in the ensemble defined by this model, and that correspond to the lowest ``energy'' characterizing the model. In particular, in Ref.\cite{Olbrich} the authors consider the simple ERG model with $6$ nodes, where only the interactions between triangles and $3$-chains are taken into account, that is, only a subset of the family of all graphs with $6$ nodes are considered. Then, the convex hull of all the possible expectation values of the probabilities of the triangles and of the $3$-chains is derived. Those graphs that correspond to the minimal ``energy'' are found to lie on the lower boundary of the mentioned convex hull. 

The geometric entropy proposed in the present work has been computed to provide  a ``pointwise'' description of the complexity of the single members of a given family of graphs. The outcomes of these computations allow to rank the members of a given family of graphs according to their degree of complexity, of course on the basis of the proposed 
way of measuring it. The results are summarized into Table \eqref{ERG}. They suggest that going up along the lower boundary of the previously mentioned convex hull (that is moving along a line representing a given family) the degree of complexity increases. 
 
 \tikzstyle{every node}=[circle, draw, fill=black!50,
                         inner sep=0pt, minimum width=4pt]
 
\begin{table}
[ht] \caption{The value  of $\widetilde{S}$ for different Exponential Random Graphs corresponding to Minimal Energy} \label{ERG}
\vspace{0.2cm}

\begin{tabular}{|clll|c|}
\hline ERG & & & &  $\widetilde{S}$\\
\hline

 \begin{tikzpicture}[thick,scale=0.8]%
    \draw \foreach \x in {30} {
        (\x:2) node{} -- (\x+60:2) node{}
        (\x:2) -- (\x:0) node{}
       (\x+60:2) -- (\x:0)
       
       (\x+17:2) node{}
       (\x+17:2) -- (\x:4)
       (\x+17:2) -- (\x+30:4)
       (\x:4) node{} -- (\x+30:4) node{}
      
};
\end{tikzpicture} & & & & $0.568$\\
\hline
\end{tabular}
\begin{tabular}{|clll|c|}
\hline ERG & & & &  $\widetilde{S}$\\
\hline

\begin{tikzpicture}[thick,scale=0.8] %
    \draw \foreach \x in {0,120} {
        (\x:2) node{} -- (\x+120:2) -- (\x+240:2) node{}
        (\x:2) -- (\x:0) node{}
        (\x+240:2) -- (\x:0)
       (\x+90:2) node{}

};
\end{tikzpicture} & & & & $1.006$\\
\hline 
\end{tabular}
\begin{tabular}{|clll|c|}
\hline ERG & & & &  $\widetilde{S}$\\
\hline

\begin{tikzpicture}[thick,scale=0.8] %
    \draw \foreach \x in {0} {
        (\x:2) node{} -- (\x+72:2) node{} -- (\x+144:2) node{} -- (\x+216:2) node{} -- (\x+288:2) node{}  -- (\x+360:2) node{}
        (\x:2) -- (\x+144:2)
      (\x:2) -- (\x+216:2)
      (\x+72:2) -- (\x+216:2)
      (\x+72:2) -- (\x+288:2)
      (\x+144:2) -- (\x+288:2)
      (\x+144:2) -- (\x+360:2)
      (\x+35:2) node{}
};
\end{tikzpicture} & & & & $1.303$\\
\hline
\end{tabular}
\end{table}

\begin{table}
[ht] \caption{The value  of $\widetilde{S}$ for Exponential Random Graphs corresponding to Maximal Energy}  \label{ERG1}
\vspace{0.2cm}

\begin{tabular}{|clll|c|}
\hline ERG & & & &  $\widetilde{S}$\\
\hline

 \begin{tikzpicture}[thick,scale=0.8]
     \draw \foreach \x in {0,60,120,180,240,300}
     {
         (\x:2) node {}  -- (\x+60:2) 
         (\x:2) -- (\x:0)

     };
 \end{tikzpicture} & & & & $2.332$\\
\hline
\end{tabular}
\end{table}

Moreover, the result of Table \eqref{ERG1} shows that our entropic measure is capable of distinguishing among different families of networks. In fact, while the graphs of Table \eqref{ERG} represent typical graphs on the minimal energy boundary, the graph of the Table \eqref{ERG1} is a typical graph on the maximal energy boundary.  
Notice that the results in Table \ref{ERG} indicate that  the network with two triangles (2-simplices) is less complex than the network with one tetrahedron (3-simplex) plus two points (0-simplices), which is less complex than the network with one 4-simplex plus one point (0-simplex); in other words, network complexity is  nontrivially influenced by network topology (homology). A first account of this fact is given in Ref.\cite{jmp}.
\bigskip

\subsection{Configuration Model}

Real networks usually differ from the Erd\"{o}s-R\'enyi random graphs in their degree distribution \cite{CNet2}. Given an undirected network with adjacency matrix $A=(A_{ij})$, the degree of a node $i$ is just the sum of the $i$-th row's entries, $d_i:=\sum_{i}A_{ij}$. It represents the number of connections that the node $i$ has. The degree distribution $P(d)$ of a network is then defined to be the fraction of nodes in the network with degree $d$. The degree distribution clearly captures information about the structure of a network. For example, in the binomial Erd\"{o}s-R\'enyi random graphs, usually indicated as simple (not-complex) networks, one finds that most nodes in the network have similar degrees; this model, in which each of the $n$ nodes is connected with independent probability $p$, has a binomial distribution of degrees $d$, namely $P(d)=\binom{n-1}{d}p^d(1-p)^{n-1-d}$ \cite{Lucz}. However, real world networks usually have very different degree distributions. That is, most of the nodes have a relatively small degree (low connectivity), while a few of them have a very large degree (i.e. are connected to many other nodes). These large-degree nodes are often referred to as hubs \cite{CNet2}.

A first step toward testing the effectiveness of our geometric entropy in quantifying the complexity of real networks is to compare networks where each node has the same given degree $d$ to networks containing hubs. When each node of a network has the same degree $d$, the network is called a $d$-regular graph \cite{Lucz}. One of the most widely used method to generate these special networks is the Configuration Model \cite{bender}. This is specified in terms of a  sequence of degrees; for a network of $n$ nodes we have a desired degree sequence $(d_1,\ldots,d_n)$, which specifies the degree $d_i$ of each node $i$, for $i=1,\ldots,n$. 

The average vertex degree $\langle d_i\rangle$ is the ratio between the total number of links in a given network and the number of nodes. It represents a first level of characterization of the topological complexity \cite{bonchev}. We consider it as benchmark to strengthen the validation of our proposal. We have numerically computed  the entropy $\widetilde{S}$ given by Eq.\eqref{collapseplot} for networks of number of nodes $n=50$, constructed as {\it random} $d$-regular graphs of two different values of $d$, that is, $d=2$ and $d=6$. A random $d$-regular graph is a random graph with the uniform distribution over all $d$-regular graphs. The computed value of the geometric entropy $\widetilde{S}$ increases with $d$, as is reported in Table \ref{uniform}. This result is very good because it is in agreement with the obvious fact that the larger $d$ the more complex the network.

\begin{table}\caption{The value of $\widetilde{S}$ for random $d$-regular graphs.}\label{uniform}
\begin{tabular}{|cc|cc|cc|}
\hline
\hskip 1truecm $n$ & & \hskip 1truecm $d$ & & & \hskip 1truecm $\widetilde{S}$\hskip 1truecm \\
\hline
\hline
\hskip 1truecm $50$ & &  \hskip 1truecm $2$ &  & & \hskip 1truecm $1.0265$\hskip 1truecm \\
\hline
\hskip 1truecm $50$ & &  \hskip 1truecm $6$ & & & \hskip 1truecm $3.8498$\hskip 1truecm \\ 
\hline
\end{tabular}
\end{table}

The next step toward real networks, consists of considering random graphs, again with a number of nodes $n=50$, and with a given sequence $d=(d_1,d_2,\ldots,d_n)$ of non-increasing degrees: $d_1\geq d_2\geq d_3\geq\ldots\geq d_n$. In so doing we proceed with the validation of the geometric entropy $\widetilde{S}$ in Eq.\eqref{collapseplot} by considering networks with one or more hubs. In the previous notation, a network with hubs is identified by one or more values in the sequence $d=(d_1,d_2,\ldots,d_n)$ which are larger than the other ones. In Table \ref{nonuniform}, the numerically obtained values of $\widetilde{S}$ are reported for networks with hubs of degree $d=8$, $d=10$, and $d=14$ respectively, while the other nodes have degree $d=2$. It is found that the complexity of a network increases with the number of hubs in it.  Moreover, Table \ref{nonuniform} shows that the network with degrees $(8,2,\ldots,2)$ is less complex than the network with degrees $(10,2,\ldots,2)$, which is less complex than the network with degrees $(14,2,\ldots,2)$; as well, the network with degrees $(8,8,2,\ldots,2)$ is less complex than the network with degrees $(10,10,2,\ldots,2)$. Again, this confirms that the geometric entropy $\widetilde{S}$ in Eq.\eqref{collapseplot} leads to an overall consistent scenario.

\begin{table}\caption{The value of $\widetilde{S}$ for networks with hubs.}\label{nonuniform}
\begin{tabular}{|cc|cc|}
\hline
$d$ & & & $\widetilde{S}$\\
\hline
\hline
$(2,\ldots,2)$ & & & $1.0265$\\
\hline
$(8,2,\ldots,2)$ & & & $1.6140$\\ 
\hline
$(8,8,2,\ldots,2)$ & & & $2.1263$\\ 
\hline
$(8,8,8,2,\ldots,2)$ & & & $2.2120$\\ 
\hline
$(8,8,8,8,2,\ldots,2)$ & & & $2.8473$\\ 
\hline
$(8,8,8,8,8,2,\ldots,2)$ & & & $3.2298$\\ 
\hline
\end{tabular}
\quad
\begin{tabular}{|cc|cc|}
\hline
$d$ & & & $\widetilde{S}$\\
\hline
\hline
$(2,\ldots,2)$ & & & $1.0265$\\
\hline
$(10,2,\ldots,2)$ & & & $1.9156$\\ 
\hline
$(10,10,2,\ldots,2)$ & & & $2.3878$\\ 
\hline
\end{tabular}
\quad
\begin{tabular}{|cc|cc|}
\hline
$d$ & & & $\widetilde{S}$\\
\hline
\hline
$(2,\ldots,2)$ & & & $1.0265$\\
\hline
$(14,2,\ldots,2)$ & & & $2.7631$\\ 
\hline
\end{tabular}
\end{table}

{ In a way, homogeneity in the interaction structures entails topological equivalence of almost all the nodes; to the contrary, it was found that most of real networks show degree distribution $P(d)=A\ d^{-\gamma}$, where $A$ is a positive \textit{real} constant and the exponent $\gamma$ varies in the range $2\leq \gamma\leq 3$ \cite{Latora}. These networks are called \textit{scale--free networks}, because power--laws have the property of self-similarity at all scales \cite{Watts,Barabasi}. Before tackling the study of scale--free networks, we now deeply focus on the issue of heterogeneity. In particular, we consider networks of the same average degree but of different degree of heterogeneity.} Understanding a regular graph as the less heterogeneous network, we see heterogeneity as a measure of how far from $P(d_0)=1,\ P(d)=1(d_0\neq d)$ a network is. This entails that the most heterogeneous network corresponds to $P(1)=P(2)=\ldots P(n-1)$ \cite{Wu}. In particular, we consider networks with $n=50$ nodes and the same average degree $d=\frac{53}{50}$; moreover, we assume the following degree sequences, $(8,2,\ldots,2)$, $(7,3,2,\ldots,2)$, $(6,4,2,\ldots,2)$, $(5,5,2,\ldots,2)$ and $(6,3,3,2\ldots,2)$. Next, the degree of heterogeneity of such networks is measured through the definition given in \cite{Estrada}, that is
\begin{equation}\label{heterog}
h=\sum_{i,j\in E}\left(\frac{1}{\sqrt{d_i}}-\frac{1}{\sqrt{d_j}}\right)^2,
\end{equation}
where $E$ is the set of edges of the network. This heterogeneity measure vanishes for regular graphs, while, as the difference in the degrees of adjacent nodes increases, it also increases. Another advantage of $h$ is that it can be expressed in terms of the Laplacian matrix \cite{Bernstein}, which is widely employed in the study of networks complexity. Therefore, to perform a benchmarking we compute the geometric entropy $\widetilde{S}$ in \eqref{collapseplot} and compare it to the heterogeneity $h$ in \eqref{heterog} for the above mentioned networks. In Table \ref{heterogeneity} we can see that the way of ordering these networks according to their decreasing degree of heterogeneity as measured by $\widetilde{S}$ is the same of the ordering produced by the $h$ measure of heterogeneity. 

\begin{table}\caption{The value of $\widetilde{S}$ for networks with same average degree and different degree of heterogeneity.}\label{heterogeneity}
\begin{tabular}{|cc|cc|cc|}
\hline
$d$ & & & $\widetilde{S}$& & \textit{h}\\
\hline
\hline
$(8,2,\ldots,2)$ & & & $1.6140$& &1\\ 
\hline
$(7,3,2,\ldots,2)$ & & & $1.3070$& & 0.8088\\ 
\hline
$(6,4,2,\ldots,2)$ & & & $1.0941$ & &0.7074\\ 
\hline
$(5,5,2,\ldots,2)$ & & & $1.0924$& &0.6754\\ 
\hline
$(6,3,3,2,\ldots,2)$ & & & $1.0357$& &0.4970\\ 
\hline
\end{tabular}
\end{table}
 
{ \subsection{Scale--Free Networks}

Many real networks show power--law degree distribution and very often they are modelled by scale--free networks, for instance this is the case of some social networks \cite{Quax13} or biological networks \cite{Tuszy}, to quote some of them.  These graphs with power--law degree distribution can be obtained as special cases of RG{\it s} with a given degree distribution, as discussed in the previous section. 

Power--laws have a particular role in statistical physics especially because of their connections to fractals \cite{Falcon} and phase transitions \cite{Stanley}. In this section, we refer to scale--free networks as the class of networks with any power--law of the degree distribution and address our investigation to the issue of phase transitions.

Beyond the two well-known RG models (uniform and Poissonian), another kind has been considered, which is suitable for describing power--law degree family. This is described by two parameters $\alpha$ and $\gamma$ which define the size and the density of a network; hence, given the number of nodes $N$ with degree $d$, such a model, denoted as $G_{\alpha,\gamma}$ \cite{Latora}, assigns uniform probability to all graphs with $N=e^{\alpha}\ d^{-\gamma}$.

The connectivity properties of the model $G_{\alpha,\gamma}$, as function of the power $\gamma$, have been shown to hold almost surely for sufficiently large graphs. Likewise, as we have previously seen, the transitional properties of the uniform random graph model $\mathbb{G}(n,k)$  hold almost surely asymptotically in $n$. For the model $G_{\alpha,\gamma}$ a critical value of $\gamma$ exists for the emergence of a giant component. It has been proved that for values of $\gamma$ smaller than $\gamma_c=3.4785$ ($\gamma<\gamma_c$) a giant component always exists. On the other hand, when $\gamma>\gamma_c$ the graphs $G_{\alpha,\gamma}$ almost surely have no giant component \cite{Aiello}.

In order to check whether $\widetilde{S}$ in Eq. \eqref{collapseplot} detects this property of scale-free networks, we have proceeded as follows. 
We considered networks of $n=250$  nodes \cite{Rick} for which, without loss of generality, we set $\alpha=0$. 
For each value of  $\gamma$, we selected $10$ different realisations of the networks, each realisation having the same value of  $k/n$ larger than the threshold value ($0.5$) for the appearance of a giant component. Actually, because of the practical difficulty of getting different realisations of a scale-free network with exactly the same value of $k/n$ at different $\gamma$, we accepted a spread of values in the range $0.7 - 0.85$.

In Fig. \ref{bifurcation2} we report the behaviour of the geometric entropy $\widetilde{S}$ of the power--law random graphs $G_{\alpha,\gamma} (n, k)$ when $\alpha=0$, $n=250$,  $r = 0.2$ (the weights of the links), and the exponent $\gamma$ is in the range $1.5<\gamma<5.5$. 

The pattern of $\widetilde{S}(\gamma)$ displays the typical phenomenon found in numerical investigations of second order phase transitions: what asymptotically would be a sharp bifurcation is here rounded because of a finite size effect. 

This is in excellent agreement with the $n$-asymptotic bifurcation at $\gamma=\gamma_c=3.47875$ predicted by the Molloy Reed criterium for the emergence of a giant component in the power--law random graphs $G_{\alpha,\gamma}$ \cite{Aiello}.

\begin{figure}[h!] \centering
\includegraphics[width=8cm,scale=1.]{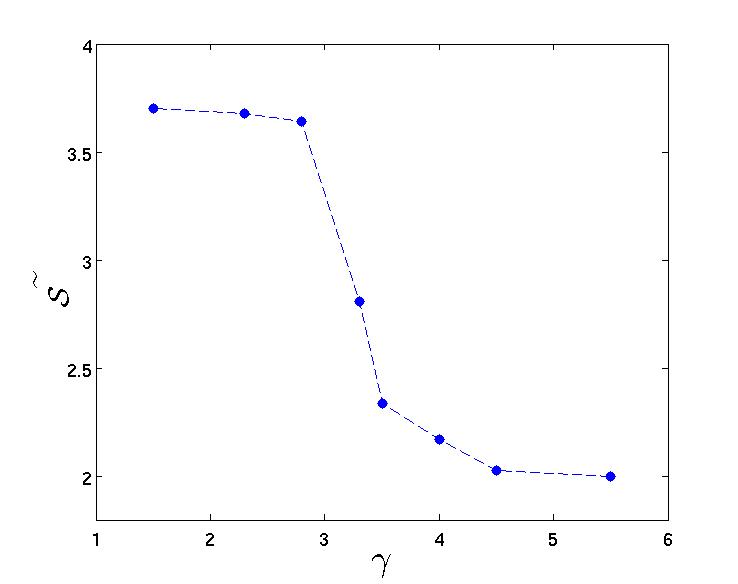}
\caption{Values of the entropy $\widetilde{\cal S}$ of power--law $G_{0,\gamma}(250,k)$ networks as a function of the exponent $\gamma$; $k$ values varied - according to the realisation of the RG and indendently of $\gamma$ -  approximately in the range $180 - 210$.}
\label{bifurcation2}
\end{figure}

}

\subsection{Real networks}

Real--world graphs are usually more complex than random graphs \cite{Watts,Barabasi}. In contrast to Erd\"{o}s-R\'enyi graphs real-world graphs have some typical features, such as, for example, power-law degree distribution, correlation of node degrees \cite{Newman02}, modularity structures \cite{Newman06}. Many complexity measures have been proposed to describe real networks capturing one or another of their typical features. Given undirected graphs $\mathbb{G}(n,k)$ with $n$ nodes and $k$ edges, some measures  { maximize the complexity of graphs with nearly the complete number of edges; other measures indicate as highly complex graphs real networks with modular structures at different levels which are expected only for a medium number of edges \cite{Wilhelm}.}  { These latter complexity measures have been basically defined  by the specification of which subnetworks are considered and when are two subnetworks different. They are outlined through (i) the one--edge--deleted subnetwork with respect the number of spanning trees, (ii) the one--edge--deleted subnetwork with respect to different spectra of Laplacian(degree matrix minus adjacency matrix) and the signless Laplacian matrix(degree matrix plus adjacency matrix) and (iii)the two--edges--deleted subgraph complexity with respect to different spectra of the Laplacian and signless Laplacian matrix. However, only the first one, denoted by $C_{1 e,\text{\tiny ST}}$ in \cite{Wilhelm}, has been applied to $33$ real networks as the other measures required a too high computational effort. Thus, the measure $C_{1 e,\text{\tiny ST}}$ has been compared to some product measures and some entropic measures.} Entropic measures quantify the diversity of different topological features { and our measure $\widetilde{S}$ of complexity may be ascribed  to this class.} { Here we compare our measure to $C_{1 e,\text{\tiny ST}}$ and discuss about the two entropic measures employed in \cite{Wilhelm}: the spanning tree sensitivity (STS) and the spanning tree sensitivity differences (STSD). These are based on the idea that complex graphs have very diverse edge sensitivities with respect to removal of different edges, while in very simple graphs all edges play the same role and the graph has the same edge sensitivity with respect to removal of different edges.}

In Table \ref{social} the outcomes of the numerical computation of $\widetilde{S}$ are reported for some of the networks considered in \cite{Wilhelm}, for which also the corresponding { $C_{1 e,\text{\tiny ST}}$} values are displayed. These networks are:  the coauthorship network of scientists working on network theory (Net Science), the coappearance network of characters in the novel \textit{Les Miserables}, the network (Dolphins) of frequent associations between dolphins, and the adjacency network (Word Net) of common adjectives and nouns in the novel \textit{David Copperfield}. Though our geometric entropy has already proved above its own meaningfulness, it is very interesting to notice that the way of ordering these networks according to their complexity which is established by $\widetilde{S}$ is the same of the ordering produced by the { $C_{1 e,\text{\tiny ST}}$} measure of complexity.  It is worth mentioning that  the network "The Miserables" is the only weighted network among those considered here, and its adjacency matrix has a relatively small number of large-weight edges. The corresponding entropy value $\widetilde{S}=1.670$, reported in Table \ref{social},  increases to $\widetilde{S}=2.644$ by setting all the weights of the edges of the network equal to $1$. Loosely speaking, this amounts to increasing the effective network connectivity, and this is correctly detected by a corresponding increase of $\widetilde{S}$.
Finally, let us note that the relative variations of $\widetilde{S}$ are much larger than those of the { $C_{1 e,\text{\tiny ST}}$} (and of other parameters defined in Ref.\cite{Wilhelm} for the same networks). This means that $\widetilde{S}$ has a greater "resolving power" in comparatively measuring the complexity of different networks. 

\begin{table}\caption{ The value of $\widetilde{S}$ for real networks: $n$ is the number of nodes, $k$ is the number of links (data taken from
{\sf http://www-personal.umich.edu/ $\widetilde{}$ mejn/netdata/} ).}\label{social}
\begin{tabular}{|c|c|c|c|c|}
\hline
\hskip 0.7truecm Network & \hskip 0.7truecm $n$ & \hskip 0.7truecm $k$ & \hskip 0.7truecm $\widetilde{S}$ & \hskip 0.5truecm { $C_{1 e,\text{\tiny ST}}$}\\
\hline
\hline
\hskip 0.7truecm Net Science & \hskip 0.7truecm $413$ & \hskip 0.7truecm $948$ & \hskip 0.7truecm $1.376$ & \hskip 0.5truecm $0.01$\\
\hline
\hskip 0.7truecm Les Miserables  & \hskip 0.7truecm $77$ & \hskip 0.7truecm $254$ & \hskip 0.7truecm $1.670$& \hskip 0.5truecm $0.10$\\ 
\hline
\hskip 0.7truecm Dolphins & \hskip 0.7truecm $62$ & \hskip 0.7truecm $159$ & \hskip 0.7truecm $2.852$ & \hskip 0.5truecm $0.14$\\ 
\hline
\hskip 0.7truecm Word Net & \hskip 0.7truecm $112$ & \hskip 0.7truecm $425$ & \hskip 0.7truecm $3.010$ & \hskip 0.5truecm $0.15$\\ 
\hline
\end{tabular}
\end{table}

{ The two entropic measures, STS and STSD, are assessed to be strongly correlated -- in terms of Pearson correlations -- to the $C_{1 e,\text{\tiny ST}}$ measure of complexity. All these measures are able to discriminate between different networks with same number of nodes and of links. However, if on the one hand, STS and STSD are performing better in characterizing the complexity of non--tree graphs, on the other hand, $C_{1 e,\text{\tiny ST}}$ best discriminates different complexities of different trees. Furthermore, STS and STSD are still slightly different. Indeed, one of them (STS) identifies more homogeneity with more complexity; on the contrary, STSD quantifies as more complex a network with a more heterogeneity \cite{Wilhelm}.}

{ Beyond the properties of assigning highest complexity values only to networks with a medium number of edges, and discriminating at best among different networks with the same number of nodes and links, a measure of complexity should assign higher values to real networks than to their randomized counterparts \cite{CNet2}.}

In Table \ref{random} the outcomes of the numerical computation of $\widetilde{S}$ are reported for the above mentioned \textit{real} networks, which are now randomized, that is, for the same number $n$  of nodes and number $k$ of links of each network, a random graph is generated. Then, the geometric entropy $\widetilde{S}$ is computed for each of these Random Graphs (RG). The remarkable result is that for each network its corresponding RG has a lower degree of complexity according to $\widetilde{S}$. 

{ 
A comparison of our measure $\widetilde{S}$ of complexity with respect to the two above mentioned entropic measures, shows that it orders the real networks of the Table \ref{social} in the same way of STS. On the other hand, it mismatches STSD as far as the 'Dolphins' and the 'Word net' networks are concerned. However, our measure assesses heterogeneous networks as more complex than homogeneous ones, exactly as the measure STSD of complexity. Furthermore, while random graphs have always higher STS values than the corresponding real networks, the STSD measure sometimes qualifies real networks as more complex than their corresponding randomized version \cite{Wilhelm}.
}

{
Summarizing, we have compared our measure $\widetilde{S}$ of complexity to the $C_{1 e,\text{\tiny ST}}$ measure which is used to quantify the complexity of a given network according to the many different subgraphs it may contain. The two different measures order the real networks listed in Tab. \ref{social} in the same way. Furthermore, such a way of ordering these real networks coincides with the STS ordering, whereas mismatches the STSD values as far as the 'Dolphins' and 'Word Net' networks are concerned. However, the fact that the randomized versions of all the real networks considered in \cite{Wilhelm} correspond to higher STS values with respect to the original networks constitutes a difficulty of this measure. At variance, the absence of such a difficulty when our geometric entropy is used, at least for the considered cases, lends further support to the validity and consistency of the measure $\widetilde{S}$. All in all, in spite of the minor discrepancies between STSD and $\widetilde{S}$ about the ordering of two real networks, both measures attribute a higher degree of complexity to networks with a higher degree of heterogeneity. Once again, all these reasons confirm that the complexity measure $\widetilde{S}$ in Eq. \eqref{collapseplot} leads to an overall consistent scenario.
}

\begin{table}\caption{ The value of $\widetilde{S}$ for real networks and their randomized: $n$ is the number of nodes, $k$ is the number of links.}\label{random}
\begin{tabular}{|c|c|c|c|c|}
\hline
\hskip 0.7truecm Network & \hskip 0.7truecm $n$ & \hskip 0.7truecm $k$ & \hskip 0.7truecm $\widetilde{S}$ & $\widetilde{S}$ of RG\\
\hline
\hline
\hskip 0.7truecm Net Science & \hskip 0.7truecm $413$ & \hskip 0.7truecm $948$ & \hskip 0.7truecm $1.376$ & $0.4454$\\
\hline
\hskip 0.7truecm Les Miserables  & \hskip 0.7truecm $77$ & \hskip 0.7truecm $254$ & \hskip 0.7truecm $1.670$&  $1.6655$\\ 
\hline
\hskip 0.7truecm Dolphins & \hskip 0.7truecm $62$ & \hskip 0.7truecm $159$ & \hskip 0.7truecm $2.852$ & $1.7246$\\ 
\hline
\hskip 0.7truecm Word Net & \hskip 0.7truecm $112$ & \hskip 0.7truecm $425$ & \hskip 0.7truecm $3.010$ & $1.4537$\\ 
\hline
\end{tabular}
\end{table}


\section{Outlook on future developments}

The geometric framework so far put forward  paves the way to interesting developments.  A relevant generalization made possible by the Riemannian geometric framework  consists in considering the time evolution of a network. In order to do this, one should drop the simplifying assumptions of the present work by adding to the $\theta^i$s of the diagonal covariance matrix $C$ also the entries $\sigma^{ij}{ :=A_{ij}}$ of the adjacency matrix $A$ as local coordinates of the statistical manifold $\widetilde{\cal M}$ of { Eqs.\eqref{varyspace} and \eqref{gvary}}. In this way the
dimension of $\widetilde{\cal M}$ increases from $n$ to $n(n+1)/2$.

Denoting with $\zeta^i=(\psi_C (A))_{lm}$ the $n(n+1)/2$  local coordinates of $\widetilde{\cal M}$, where
$i = \sum^{l-2}_{r=0} (n - r) + m - l + 1$, there is a natural way of tackling the dynamical evolution of the network associated with $(\widetilde{\cal M},\widetilde{g})$, that is,  through the geodesic flow  given by the following set of equations
\begin{equation}
\dfrac{d^2 \zeta^i}{d s^2} + { \sum_{jk}}\;\Gamma^i_{jk} \dfrac{d \zeta^j}{d s} \dfrac{d \zeta^k}{d s} = 0 \, 
\hskip 0.7truecm i,j,k=1,\dots,n(n+1)/2
\label{eqgeodesic}
\end{equation}
where the $\Gamma^i_{jk}$ are the
standard Christoffel connection coefficients 
\begin{equation}
\Gamma^i_{jk} = \frac{1}{2} { \sum_{l}}\; \widetilde{g}^{il} \left(
\partial_j \widetilde{g}_{lk} + \partial_k \widetilde{g}_{jl} -\partial_l \widetilde{g}_{jk} 
\right) \, .
\label{christoffel}
\end{equation}

The physical time parametrization of the arc length $s$ is derived by means of the metric tensor as
\begin{equation}
\dfrac{d s}{dt} = \sqrt{{ \sum_{ij}}\; \widetilde{g}_{ij} \dot{\zeta}^i \dot{\zeta}^j} \, ,
\label{propertime}
\end{equation} 
where the $\dot{\zeta}^i$ are the variation rates of the local coordinates expressed with respect to the physical time $t$. Let us remark that the dynamical evolution described by Eq.\eqref{eqgeodesic} encompasses also the time evolution of the weights of the links of a network, including their appearance and disappearance,   thus a-priori allowing the computation of the time variation $\widetilde{\cal S}(t)$ of its complexity. The  fitting of empirical data concerning the true evolution of a real network by means of the model dynamics given  by Eq.s \eqref{eqgeodesic} and \eqref{propertime} could allow to get relevant information about the laws that drive the network evolution (conservation, extremalization, optimisation of some quantities and so on).

Another prospective and remarkable application of the differential geometrical approach put forward in the present work, and notably related with the dynamical equations \eqref{eqgeodesic},  concerns the study of the stability properties of a network. In fact, by setting $\zeta^i (t)\to \zeta^i(t) + \varphi^i(t)$, where $\varphi^i(t)$ are small functional perturbations, after substitution into Eq.\eqref{eqgeodesic} and using  \eqref{propertime} one can work out the \textit{tangent dynamics equations} in the form of a system of first order linear differential equations \cite{P07}
\begin{equation}
\frac{d\varphi^i}{dt} = \Phi^i(\{\zeta^j\})\ .
\label{tgdyn}
\end{equation}
These equations, numerically integrated together Eqs. \eqref{eqgeodesic} and \eqref{propertime}, are the natural tool to investigate the stability of either stationary or non stationary states of a network, for example -  for a stationary state - 
to investigate a network stability/instability under addition or deletion of one or more links and so on.

\section{Conclusion}

Summarizing, the present work contributes the fascinating subject of quantifying the degree of complexity of systems that are commonly defined as ``complex".
There is a large number of such definitions that are already available. Perhaps this history begins with Kolmogorov's definition of algorithmic complexity \cite{kolmogo,kolmogo1} which, in spite of its theoretical beauty, is hardly applicable in practice. Since then the many definitions put forward were adapted to the specific systems/problems tackled. However, the number of categories in which all these definitions can be gathered is rather limited. Of course borrowing the concept of physical entropy from statistical mechanics is the most inspiring and seducing way to proceed. In fact, in physics, entropy is just a measure of disorder and conversely negentropy - as defined by L.Brillouin a long time ago - is a measure of the degree of order in a system and is also the physical equivalent of Shannon's information entropy, again, as shown by L.Brillouin \cite{brillouin}. Whence
a vast literature addressing the quantification of complexity  on the basis of Shannon's information entropy
which, on the other hand, has its inspiring model in Boltzmann's entropy of kinetic theory. However, what was still missing was a general definition of an entropic measure of complexity accounting for both the structure of any given network and for its statistical complexity, that is, for the complexity of the probability distributions of the entities constituting the network. The new definition put forward in the present work embraces both these aspects. It is still inspired to statistical mechanics, however, instead of being modeled on Boltzmann entropy is rather modeled on the microcanonical ensemble definition of entropy. The phase space volume being replaced by the volume of a ``state manifold" (that is a Riemannian manifold whose points correspond to all the possible states of a given network). The state manifold is defined through a suitable definition of a metric which is partly borrowed from the so-called Information Geometry and partly is an original proposal put forward in the present work. The result is a constructive way of associating a differentiable and handy mathematical object to any simple undirected and weighted graph or network. Another novelty consists in having directly tested by means of numerical computations the validity and effectiveness of the proposed entropic-geometric measure of complexity. In order to do this we needed, so to speak, a paradigmatic example of a major change of complexity. A possible natural choice is suggested by the observation that phase transitions are the most impressive examples in nature of emergent phenomena - theoretically well understood - associated with a sharp qualitative and quantitative change of complexity of a physical system when a control parameter exceeds a critical value. This kind of phenomenon exists also in complex networks. In fact, random graphs undergo a well known phase transition as proved by the Erd\"{o}s-R\'enyi theorem:  a paradigmatic - and at present unique - example of an analytically known major variation of the degree of complexity of a network. This kind of check is unprecedented and very successful, in fact,  the entropic-geometric measure of complexity proposed here displays both a pattern and its size-dependence which are typically found for the order parameter of a second-order phase transition in physics.
Then, since the random graphs undergoing the Erd\"{o}s-R\'enyi transition are not considered genuinely complex networks, the proposed entropic-geometric measure of complexity has been applied to small Exponential Random 
Graphs, to different versions of random $d$-regular graphs with and without hubs generated according to the Configuration Model. { Moreover, the ability of detecting the transitions predicted by Molloy Reed criterium in power--law random graphs is an important confirmation of the consistency of the geometric entropy measure of complexity proposed in the present. Finally, this has been applied to some real networks and compared to three other measures of complexity, one describing the modularity structures, one homogeneity, and the heterogeneity of the real networks.} The outcomes of these applications compose a consistent scenario validating the meaningfulness and effectiveness of the proposed measure of complexity.

Finally, the differential-geometric framework put forward opens some fascinating perspectives of application  to the study of the time evolution of complex systems.

\begin{acknowledgments}
We are indebited with R. Quax for providing us with data for power-law RG. We also thank M. Rasetti for useful discussions. This work was supported by the Seventh Framework Programme for Research of the European Commission under FET-Proactive grant  TOPDRIM (FP7-ICT-318121).
\end{acknowledgments}

\end{document}